\documentclass[review]{elsarticle}

\usepackage{hyperref}

\usepackage{multirow}
\usepackage{color}
\usepackage[dvipsnames]{xcolor}

\journal{Journal of Information Processing and Management}

\begin{document}

\begin{frontmatter}

\title{Newswire versus Social Media for Disaster Response and Recovery 
}


\author[mymainaddress]{Rakesh Verma\corref{mycorrespondingauthor}}
\author[mysecondaryaddress,mymainaddress]{Samaneh Karimi}
\author[mymainaddress]{Daniel Lee}
\author[mymainaddress]{Omprakash Gnawali}
\author[mysecondaryaddress,mythirdaddress]{Azadeh Shakery}

\cortext[mycorrespondingauthor]{This is to indicate the corresponding author.}

\address[mymainaddress]{University of Houston}
\address[mysecondaryaddress]{University of Tehran}
\address[mythirdaddress]{Institute for Research in Fundamental Sciences (IPM)}
\begin{abstract}
In a disaster situation, first responders need to quickly acquire situational awareness and prioritize response based on the need, resources available and impact. Can they do this based on digital media such as Twitter alone, or newswire alone, or some combination of the two? We examine this question in the context of the 2015 Nepal Earthquakes. Because newswire articles are longer, effective summaries can be helpful in saving time yet giving key content. We evaluate the effectiveness of several unsupervised summarization techniques  in capturing key content. We propose a method to link tweets written by the public and newswire articles, so that we can compare their key characteristics: timeliness, whether tweets appear earlier than their corresponding news articles, and content.  A novel idea is to view relevant tweets as a summary of the matching news article and evaluate these summaries. Whenever possible, we present both quantitative and qualitative evaluations. One of our main findings is that tweets and newswire articles provide complementary perspectives that form a holistic view of the disaster situation. 
\end{abstract}

\begin{keyword}
Nepal Earthquake  \sep newswire \sep social media \sep summarization \sep disaster response \sep disaster recovery
\MSC[2010] 00-01\sep  99-00
\end{keyword}

\end{frontmatter}


\section{Introduction}
When a disaster strikes, responders and relief agencies need to rapidly assess the damages to lives and infrastructures, and get a grip on the situation. Phone service and electricity supply may be disrupted in various parts of the affected region. Thus, there may not be direct sources of information available, e.g., calling (or messaging) the first responders in the affected region may not be possible. How to acquire important and reliable information quickly in such a fast-moving and chaotic situation? For this purpose, they may turn to indirect sources: the social networks, such as Twitter, or the  newswire services.  

Twitter has become a {\em de facto} standard domain for event detection~\cite{petrovic13}, because of its real-time nature. Researchers have given some evidence to show that Twitter users break news before newswire. However, few studies have examined the content and timeliness of the two different sources, especially in the context of a major disaster. 

In this paper, we examine the content and timeliness of the two different sources in the context of the 2015 Nepal Earthquakes. Since newswire articles are typically much longer than tweets, and first responders  may not have much time to read multiple news articles before responding in a disaster situation, we also examine the effectiveness of summarization from several viewpoints. These considerations lead to the following research questions (RQ):
\begin{enumerate}
    \item Does Twitter report news faster than traditional newswire, especially in the context of a rapidly changing situation such as a major disaster? (RQ1) 
    \item What type of information is reported earlier by Twitter, especially in the context of a rapidly changing situation such as a major disaster? (RQ2)
    \item How effective is summarization of news articles with respect to Twitter content? (RQ3)
    \item How effective is summarization of news articles for capturing the important information from human viewpoint? (RQ4)
    \item How effective are some of the methods proposed for linking tweets with news articles?  (RQ5)
    \item How effective are tweets as summaries of news articles? (RQ6)
\end{enumerate}
The rest of this paper is organized as follows. In the next section, we review the relevant related work. In Section~\ref{datasets}, we describe the datasets. Section~\ref{linking} describes the proposed method for linking tweets with news articles and its evaluation (RQ5). The summarization research questions (RQ3 and RQ4) are investigated in the following section. RQ1 and RQ2 are studied in Sections \ref{NewsRepSpeedSec} and \ref{tweet-events} and Section \ref{conclusion} concludes the paper. 

\section{Related Work}
Twitter for emergency applications has been studied by several researchers, e.g., \cite{millsCL09,cassaCM13,spenceLL15,doanVC11,martinez18}. In~\cite{millsCL09}, researchers concluded that Twitter was not yet ready for first responders. However, it was helpful for civilians. These were the early days of Twitter, as we find from \cite{cassaCM13} that individuals immediately posted specific information helpful to ``early recognition and characterization of emergency events'' in the case of  the Boston marathon bombing. In~\cite{spenceLL15}, researchers found that tangible, useful information was found in the early period before storm system Sandy and it got buried in emotional tweets as the storm actually hit. However, we think more studies are needed on this issue, since the tweets collected were rather small, approximately 27,000, using just the hashtag \#sandy. A bilingual analysis of tweets obtained over 84 days overlapping the Tohoku earthquake showed, among other results, the correlation between Twitter data and earthquake events~\cite{doanVC11}. A survey of this literature can be found in~\cite{martinez18}. 

Several papers have examined Twitter data in the Nepal Earthquake context ~\cite{alamJI18,hurriyetogluGO16,radiantiHL16,subbaB17,suLL16}. Relevance of tweets was examined by~\cite{alamJI18,hurriyetogluGO16}.  However, note that our problem is different, viz., whether a tweet is relevant in the context of a given news article. The other papers examined different aspects such as public concerns and perceptions of disaster recovery efforts~\cite{radiantiHL16,suLL16} and public reaction to social media project of the police~\cite{subbaB17}. 

Researchers have examined the question of whether Twitter can replace newswire for breaking news \cite{petrovic13}. They studied a period of 77 days in 2011 during which 27 events occurred. The biggest disasters in this event-set are: an airplane crash with 43 deaths, and a magnitude 5.8 earthquake in Virginia that caused infrastructural damage.\footnote{None of these disasters, bad as they are, rise to the level of the Nepal Earthquake(s) of 2015 in which almost 10,000 lives were lost.} They collected a large dataset of tweets and news articles, but then eliminated a large collection of tweets based on clustering. More elimination of tweets led to only 97 linked tweet-news article pairs, which is a small dataset.

Thus, some researchers have focused on comparing the two sources of information, e.g., \cite{petrovic13}, some others utilize the joint information in them to improve the performance of news related tasks, and some papers try to discover the linkage between tweets and news articles \cite{tsagkias2011, mogadala2017, wang2015, ahmad2016, mazoyer2018, lin2016}.
\subsection{Tweet-News Linking}
In this section, the previous works on these areas are reviewed.

 The tweet-news linking method proposed by Guo et. al, \cite{guo2013} with the aim of enriching short text data in social networks is a graph based latent variable model. They extract text-to-text relations using hashtags in tweets and named entities in news articles along with their temporal similarity.

In \cite{shi2014}, a framework for connecting news articles to Twitter conversations is proposed using Local cosine similarity, global cosine similarity, local frequency of the hashtag and global frequency of the hashtag as the classification features extracted for each article-hashtag pair.
The task of linking tweets with related news articles is studied in another paper  to construct user profiles \cite{abel2011}. The authors proposed two sets of strategies to find relevant news articles to each tweet in this paper. In addition to URL-based strategies, which is similar to the idea used in \cite{mccreadie2013}, they also proposed several content-based strategies that include computing the similarity between hashtag-based, entity-based and bag-of-word-based representations of tweets and news articles to  discover the relation between them. In addition to user modeling, the tweet-news linking task has been employed in document summarization \cite{wei2015}, sentiment analysis \cite{kulcu2016} and event extraction \cite{li2016} 

\subsection{Tweet Assisted Summarization and Tweet Summarization}
In~\cite{weiG15}, researchers proposed two methods that leverage tweets for ranking sentences in news articles for summarization:  a voting method based on tweet hit counts of sentences, and a random walk on a heterogeneous graph (HGRW) consisting of tweets and news article sentences as nodes and the edge weights are defined by weighted idf-modified-cosine scores. The best ROUGE-1 F-score~\cite{linH03} is achieved by a version of HGRW that outputs both sentences from news articles and tweets in the summary, where the summary consists of top four sentences/tweets as highlights of the article. 

Tweet summarization has also been studied, e.g., see ~\cite{rudraBG16} and references cited therein. Our problem is a little different, we consider tweets that are linked and found relevant (or partially relevant) to news articles from the perspective of summaries of those news articles. We then evaluate them to get an idea of how much content of the articles is captured by these tweets. 

\section{Datasets} \label{datasets}
We collected two datasets, a tweet dataset, and a newspaper article dataset, for our research questions. We describe these datasets below.

\subsection{Twitter Dataset \& Characteristics}\label{TwitterAnalysis}
There are some tweet datasets on the Nepal Earthquake released by other researchers, e.g., \cite{firoj_ACL_2018embaddings}. However, since the main goal in this paper is to use the linkage between tweets and news media content for disaster response and recovery, we need two contemporaneous datasets of news articles and tweets, relevant to the Nepal Earthquake. Therefore, we collected a set of tweets about the Nepal earthquake using ``Nepal earthquake'' as the search query and annotated them. 
We also collected a set of news articles as explained in section \ref{sec:newsArticleCollectionSec}.
The tweet collection consists of 336,140 tweets written from April 24, 2015 to June 25, 2015. In the rest of this section, some characteristics of the tweets generated in this time period is studied. 
\subsubsection{Most Frequent Words in Tweets}
As shown in table \ref{TweetFreqWords}, the top 10 most frequent words in the set of collected tweets are nepal, earthquake, help, relief and victims. The number of unique words in this dataset is 91,752.

\begin{table}[h]
\centering
\caption{Most frequent words in tweets}
\label{TweetFreqWords}
\begin{tabular}{ll} \hline
Word & Frequency \\ \hline
nepal & 329,836 \\ 
earthquake & 312,721 \\ 
help & 31,012 \\ 
relief & 26,062 \\ 
victims & 24,823 \\ \hline

\end{tabular}
\end{table}

\subsubsection{The Number of Tweets about the Nepal Earthquake over Time}
\begin{figure}[!htb]
\centering
\includegraphics[width=\textwidth]{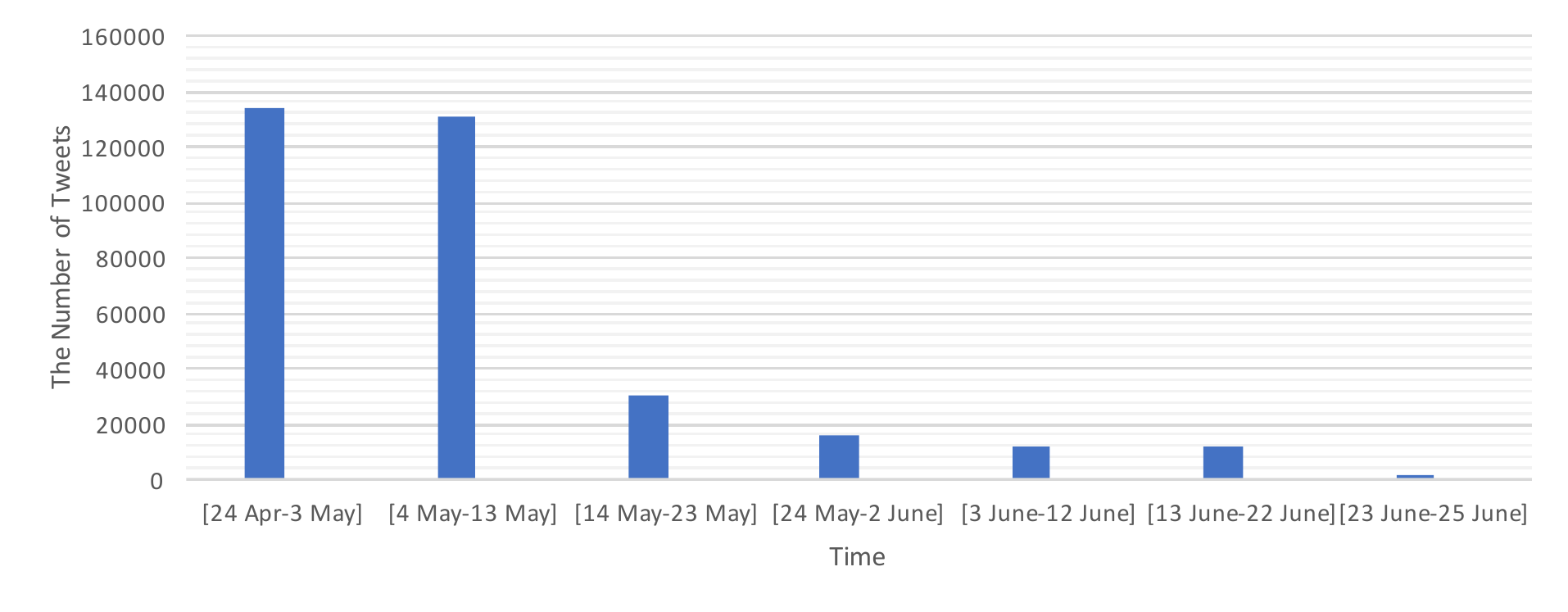}
\caption{The number of tweets written about the Nepal earthquake over time.}
\label{fig:tweetCountOverTime}
\end{figure}

The number of tweets written about the Nepal earthquake varies in different points of time as shown in Figure~\ref{fig:tweetCountOverTime}. The higher number of tweets during the first two periods, i.e. from April 24 to May 23, is due to the occurrence of the earthquake on April 25, its aftershocks and all related issues that raised consequently. 

\subsubsection{Twitter Users Activity}
The tweets of the collected tweet dataset is generated by 160,053 unique users. The top 4 users ranked based on the number of written tweets from April 24, 2015 to June 25, 2015 about the Nepal earthquake are shown in Table~\ref{TwitterUsersAcitivity}. According to their descriptions, the top ranked users are those who are active in news area in different related domains such as Nepal related news, natural disasters or Muslim world news.

\begin{table}[h]
\centering
\caption{The top 4 most active users in the Nepal earthquake tweets dataset}
\label{TwitterUsersAcitivity}

\begin{tabular}{ccc} \hline
Username & Description & Number of \\
& & Tweets \\ \hline
dlnepalnews & Link you with news in Nepal. & 1,601 \\ 
gcmcEarthquake & We tweet about Crisis, Disaster and & 1,078\\
&  Emergency Management related to & \\ & Earthquakes. &  \\ 
crowdtrendies & The latest uber great campaigns
from & 752\\
& all your favourite crowd funding websites. &  \\
wilayah\_news & Breaking news and information
from &673\\
& the Muslim world. &  \\ \hline

\end{tabular}
\end{table}

\subsubsection{The Descriptive Statistics of the Twitter Dataset}
\label{descriptiveSection}
Table \ref{DescriptiveStatTweet} shows some of the descriptive statistics of the collected tweet dataset including the fraction of tweets with URLs, mentions, hashtags and disaster keywords. To compute the fraction of tweets that contain a disaster keyword, a list of keywords about weather, disaster and emergency \footnote{\url{https://gist.github.com/jm3/2815378}} is employed.

\begin{table}[h]
\centering
\caption{Descriptive statistics of the collected Twitter dataset}
\label{DescriptiveStatTweet}
\begin{tabular}{ll} \hline
Percentage of tweets with mentions & 17.3\%\\ 
Percentage of tweets with URLs &  77.3\% \\
Percentage of tweets containing disaster  keywords & 40.6\%\\ 
Percentage of tweets with hashtags & 32.1\% \\ \hline
\end{tabular}
\end{table}

\subsubsection{Annotation}
To evaluate the performance of the proposed tweet-news linking method, a set of 310 pairs of tweet-news articles were selected to be annotated by a group of 10 researchers, such that each pair was annotated by two annotators. 
This set of 310 pairs is the result of pairing a set of 31 news articles and their top 10 most similar tweets based on their TFIDF similarity score. The set of 31 news articles is a subset of the news articles used for summarization annotation, explained in section \ref{sec:NewsAnnotation}, that are selected based on being relevant to the Nepal earthquake.
They were asked to annotate each pair of tweet-news article according to the following instructions. 
\begin{itemize}
    \item If the tweet is relevant to the news article, i.e. it is about a specific subject that is also mentioned in the news article, it should be labeled as relevant (label = 2).
    \item If the tweet is generally relevant to the topic of the news article, it should be labeled as partially relevant(label = 1).
    \item If the tweet doesn't have a meaningful content or is totally irrelevant to the news article, it should be labeled as not relevant (label = 0).
\end{itemize}
One example of each type was also included to further clarify these categories.

\subsection{News Articles Dataset \& Characteristics}
\label{sec:newsArticleCollectionSec}
 News articles were collected from five  Nepali news sources: ``Kantipur,'' ``Kathmandu Post,'' ``Nayaptika,'' ``Nepali Times'' and ``The Rising Nepal.'' In total more than 700 articles were obtained in both English and Nepali, all of them published between April 2, 2015 and November 5, 2015. The English subset ranges from April 28, 2015 to August 19, 2015.

\subsubsection{Some Challenges of the News Dataset}
For the present paper, we focus only on the English language subset of the dataset. Hence, we first filter out any files that have Nepali filenames or content. 

Because the articles were printed in Nepal, even the English language subset had encoding that was incompatible with standard English characters. So the articles were filtered further to include only those that can be decoded to ASCII characters. 

After this, another round of filtering was done to ignore articles which contained less than 1000 characters (approx. 100 words). And the final dataset used for annotation required a minimum of 10 sentences. 

\begin{table}[h]
    \centering
    \caption{Details of files remaining after filtering the News Dataset.}
    \begin{tabular}{lll}
        \hline
        Filter & Articles Remaining & Extra Info \\ \hline
         English-Only & 799 & 5 were empty files  \\
         1000 character minimum & 517 & 414 avg. words per article \\
         10 sentence minimum & 349 & 519 avg. words per article \\
         \hline
    \end{tabular}
    \label{tab:filter-details}
\end{table}

\subsubsection{Dataset of Human Annotations}
The set of 349 news articles was used to make a smaller annotated set, by dividing into articles with less than 20 sentences and those with greater than 20 sentences. Then 36 were randomly selected from the set of those less than 20 sentences, and 24 from the set more than 20 sentences. There were 12 annotators and each annotated 10 articles: 6 smaller and 4 larger. For each set of ten articles was annotated by two annotators. This resulted in a set of 60 human annotated news articles with summaries. Annotators were asked to perform two annotation tasks for each article (Figure~\ref{fig:task-prompt}): (1) abstractive and (2) extractive \footnote{Annotator H extractive summaries missing for 3 documents due to unforeseen issues.}

\begin{figure}[h]
    \centering\small
    \fbox{\begin{minipage}{0.95\linewidth}
    \setlength{\baselineskip}{0.5pt}
    ``I'm working on research related to Nepali earthquake and news. I need your help with the following task. Please read everything carefully before starting your tasks (especially the list of words in email attachment). You will receive a link to a Google sheet separately, shortly after receiving this email. The Google sheet tells you which documents you should summarize. The ID matches the text file in the zip file attached.''
    
    ``Overall task is to annotate document to be used in summary related research.
    \begin{itemize}
        \item Task \#1 In 100 words (or less) please summarize each document.
        \begin{itemize}
            \item If mentioned, your summary must include all of the following.
            \item Statistics of earthquake (i.e. Richter scale, duration).
            \item Number of aftershocks.
            \item Damages in monetary units.
            \item Deaths.
            \item Injuries.
            \item Locations
            \item \# of buildings damaged.
            \item any enumeration of damage.
            \item Also if any of the words in attached document are mentioned please include in summary.
        \end{itemize}
        \item Task \#2: Select the 5 most relevant sentences of the document. Please report the sentence numbers in order of decreasing importance. The sentence number is the position in the document (e.g. first sentence id is 1, next is 2, etc...)
    \end{itemize}
    If there is too much information to be able to summarize please mark yes in the `too much info' column of spreadsheet. But still do your best to complete both tasks.''
    \end{minipage}}
    \caption{Task Prompt for Human Annotators.}
    \label{fig:task-prompt}
\end{figure}

\begin{table}[h]
    \centering
        \caption{List of words in attachment for human annotation task.}
    \begin{tabular}{lllll}
    \hline
    Emergency & Exigency & Hurricane & Tornado & Twister \\
    Tsunami & Earthquake & Quake & Seism & Temblor  \\
    Tremor & Flood & Storm & Crest & Extreme weather\\
    Forest fire & Brush fire & Ice & Stranded & Avalanche\\
    Hail & Wildfire & Magnitude & \multicolumn{2}{l}{Shelter-in-place} \\
    Typhoon & Stuck & Disaster & Snow & Blizzard\\
    Sleet & Mud slide & Mudslide & Erosion & Power outage\\
    Brown out & Warning & Watch & Lightening & Lightning\\
    Aid & Assistance & Help & Relief & Closure\\
    Closedown & Closing & Shutdown & Interstate & Burst\\
    \multicolumn{2}{l}{Emergency Broadcast System} & \multicolumn{2}{l}{Tsunami Warning Center }\\
    \hline
    \end{tabular}
    \label{tab:attachment-word-list}
\end{table}

\subsubsection{Preprocessing}
For consistency all data analyzed in conjunction with the News Dataset used the following preprocessing steps:
\begin{itemize}
    \item News article content was parsed into sentences, and then each sentence into word tokens.
    \item Stopwords (words with low information value) were removed.
    \item Stemming was done to allow words like ``work,'' ``worked'' and ``working'' to be considered as the same.
    \item During any comparison, the word tokens were always lowercased.
\end{itemize}
The preprocessing was done using the open-source package Natural Language Toolkit (NLTK) \cite{bird2009nlp}.

\subsubsection{Annotation}
\label{sec:NewsAnnotation}
To keep the task of creating summaries feasible, 60 documents were chosen for annotation. These were annotated by a group of 12 researchers with the following instructions. Each annotator was provided the same list of keywords. 
\begin{itemize}
    \item Summarize each article in 100 words (or less).
    \item If a keyword is mentioned, then it must also be mentioned in the summary,
    \item In addition to the 100 word summary,  select five  sentences that best represent the document content.
\end{itemize}
For inter-annotation agreement tests, each document was annotated by two annotators. We separated the initial English-only subset into articles that were between 10-20 sentences long and articles with more than 20 sentences. Each set of ten documents included six from the former set and four  from the latter. The same set of ten documents were reviewed by two annotators. 

\section{Linking Tweets with News Articles} \label{linking}
To discover the linkage between tweets and news articles, a machine learning approach is proposed that explores the space of tweet-news article pairs. In this approach, each pair of tweet and news article is represented by a set of features, then a classification model is learned using a training set with matching labels. Finally, the trained model is applied on the tweet-news article pairs of the test set to find the matched ones.
The features employed to represent each tweet-news article pair are as follows.

\textbf{char Ngram similarity score}: This feature measures the similarity between the char Ngrams of the tweet and the news article by counting the number of the matched char Ngrams normalized by the total number of possible matched character Ngrams. The main aim of defining and using this feature is to detect the similarities between tweet's words, specially hashtags, written in camel case style and news article words. In this paper, this feature is computed for N = 2 and N = 3. Furthermore, the same feature is computed for the expanded versions of the tweet and news article datasets. For expansion, we add all WordNet synsets \citep{Miller1995} of each word found in the text. Thus, four different features are computed for each tweet-news article pair including char2gramSim, char3gramSim, exp\_char2gramSim and exp\_char3gramSim.

\textbf{Temporal distance}: The difference between the publish date of the news article and the tweet is considered as a feature for each pair.

\textbf{TFIDF score}: The TFIDF similarity between the tweet and news article content is employed as another feature. The TFIDF similarity scores are calculated using the Lemur project \footnote{ \url{https://www.lemurproject.org}}.

\textbf{Hashtag similarity}: 
This feature is computed by counting the number of hashtags matched with any term in the news article, normalized by the total number of hashtags used in the tweet.

After obtaining the feature vectors of the training set and test set, the classification model should be learned. Since the pairs of tweet-news articles are used as the instances in this paper and the number of matched pairs are very few compared to all possible pairs of tweets and news articles in the training set, the training data is imbalanced. Therefore, a random undersampling method is employed to make the training data balanced before learning the classification model. In this paper, SVM is used as the classification method with parameters tuned using a validation set. We randomly selected one fifth of the training set and used it as the validation set for parameter tuning. 

\subsection{Experimental Results}
\subsubsection{Training and Test Sets}
As mentioned in section \ref{TwitterAnalysis}, tweets related to the Nepal earthquake were collected. Then, the list of keywords about weather, disaster and emergency, mentioned in \ref{descriptiveSection}, is used to filter the tweets that contain at least one of the keywords. The final set of tweets is employed as the test set. We used another dataset of tweets and news articles with their matching labels \cite{Guo13} as the training set. 
The number of tweets in the training set is 34,888 and the number of news articles in the training set is 12,704. 
Both training and test sets are preprocessed by removing the non-ascii characters, punctuation, stop-words and URLs. 
We created the feature vectors for all pairs of tweets and their top (up to) 100 retrieved news articles, based on their TFIDF similarity score, for both training and test sets. In summary, the total number of instances, i.e. tweet-news pairs, in the training data is 759,971 and in the test data is 528,402. 

\subsubsection{Experiments}
\label{sssec:tweet-relevance}
 We used the Scikit-Learn and imblearn Python libraries for the implementation of the classification and undersampling methods respectively. We randomly selected one fifth of the training set and used it as the validation set for tuning gamma and C, the parameters of the SVM method. To tune the parameters, we performed 5-fold cross validation on the validations set using the GridSearchCV module of scikit-learn library. 

For each news article, we ranked the tweets based on their class membership probabilities, namely the probabilities of being classified as relevant, estimated by SVM. We selected the top 10 tweets for each news article and annotated the resulting pairs as relevant, partially relevant or irrelevant for a subset of 31 news articles from the test set.
To aggregate the two labels assigned by two annotators for each pair, we considered the ceiling of their arithmetic mean as the final annotation for that pair. In other words, if the sum of two labels is 0, the final annotation for that pair would be 0 (i.e. irrelevant pair), if the sum of two labels is either 1 or 2, the final annotation would be 1 (i.e. partially relevant pair) and if the sum of two labels is either 3 or 4, the final annotation would be 2 (i.e. relevant pair). After obtaining the aggregated annotations for each pair, the precision is calculated. To compute the precision, we considered the partially relevant examples as true positive with a weight of 0.5. The precision and aggregated annotation results on the set of 310 tweet-news pairs are shown in table \ref{TweetNewLinkingPrecision}. 

\begin{table}[h]
\centering
\caption{The precision and annotation results on the set of 310 pairs.}
\label{TweetNewLinkingPrecision}
\begin{tabular}{ccc} \hline
The number of &  The number of & Precision\\
relevant pairs & partially relevant pairs & \\ \hline
37 & 218 & 0.47 \\ 
\hline
\end{tabular}
\end{table}

We computed the actual agreement between the two annotators by considering a weight of 0.5 for which the annotations difference equals to one. The computed score for our annotations is 0.59 which means that there is a moderate agreement between the annotators.

\subsection{Challenges of the Tweet-News Linking Task}
Some of the challenges of the tweet-news linking task are:
\begin{itemize}
    \item The lack of published time information of news articles.
    \item The lack of geographic location information of tweets.
    \item Different time zones in tweets' time information.
\end{itemize}

\section{Summarization of News Articles}
Summarization is used to paraphrase or represent a large document with a smaller set of sentences. Here we look at how summarization can be used to alleviate some of the information overload of articles related to the Nepal Earthquake. We analyze the datasets used and present some results of automatically generated summaries.

\subsection{Evaluation of Annotation}
There are many ways to evaluate the annotations of humans. One simple measure is the Jaccard Index. It computes the amount of common elements between two sets. In the case of our task, a set would be a single summary and the elements would be words of that summary. Table~\ref{tab:jaccard} reports the average Jaccard Index over the ten documents annotated by each annotator pair. Note that one set of 10 documents had only a single annotator so the Jaccard Index is not applicable. However, the annotations are still used in all relevant evaluations. 
\begin{equation}
    Jaccard(A, B) = \frac{ |A \cap B| }{ |A \cup B| }
\end{equation}

\begin{table}[h]
    \centering
    \caption{Average Jaccard Index between each pair of annotators.}
    \begin{tabular}{ll}
        Annotator Pair & Jaccard Index \\
        \hline
         a and e & 0.3213 \\
         b and d & 0.1877 \\
         c and k & 0.1384 \\
         f and g & 0.2783 \\
         h and i & 0.2299 \\
         j and l & N/A \\
         \hline
    \end{tabular}

    \label{tab:jaccard}
\end{table}

\subsection{Limits of Extractive Summaries}
ROUGE will be the metric of choice for the evaluations of automatic summarizers. ROUGE is based on the word overlap between peer summary (i.e. generated summary) and an annotated summary. Since, annotator generated abstractive summaries will use words not in the vocabulary of the documents, even a perfect extractive summarizer will never achieve 100\% recall. 

We investigate this limit by using the {\em articles} themselves as a peer summary against their matching annotator summaries. Because we will be evaluating the whole article, the precision will be low. However, the recall value can show the upper limit of any extractive summarizer. Running ROUGE shows that the maximum achievable score is 89.8\%. This puts into context the ROUGE F1 scores reported earlier, and reinforces a similar result of \cite{vermaL17}, where they also computed this limit for Document Understanding Conference (DUC) data.

\subsection{Evaluating Automatic Summarizers}
\label{ssec:eval-auto-summ}
We created summaries of the English-only subset using the following approaches:
\begin{enumerate}
    \item PKUSUMSUM an open source package that has algorithms for single-document, multi-document, and topic-based summarization \cite{ZhangWW16}.
    \item DocSumm is a Python package that has greedy and dynamic programming heuristics for approximating the summary based on sentences of original document (extractive summarization)~\cite{vermaL17}. 
    \item NewsSumm is a Python package that has several heuristics, including integer linear programming (ILP) and title-based reduction of a document before summarization, for extractive summarization. 
\end{enumerate}

The summaries of the human annotated articles are then evaluated using ROUGE software \cite{lin:rouge}, since ROUGE needs reference summaries. We report the ROUGE 1-gram F1 score for summaries created by PKUSUMSUM, DocSumm, and NewsSumm in Table~\ref{tab:summarizers}. For each generated summary we compare them against the human generated summaries (abstractive) and also the human selected representative sentences (extractive). 
Since extractive summaries are based on human selection of the five best sentences,  of arbitary length, to represent the document, it would be better to evaluate ROUGE-1 without a 100-word limit on the human-annotated extractive summary. Therefore, precision is a better metric for comparison and those are the results presented in Table~\ref{tab:summarizers}. We also report ROUGE-2 scores, because research has shown that in some cases this is a better suited metric \cite{OwczarzakCDN12,Graham15}.

\begin{table}[h]
    \centering
    \caption{ROUGE Evaluation of Extractive Summarizers against Human Annotated Summaries using ROUGE Unigram (R1) and ROUGE Bigram (R2).}
    \begin{tabular}{lcc}
    Method & R1/R2 F1 & R1/R2 Prec. \\
    & Abstractive & Extractive \\
    \hline
    PKUSUMSUM - Lead & 0.58618 / 0.47536 & 0.68469 / 0.55647 \\
    PKUSUMSUM - Centroid & 0.42837 / 0.20766 & 0.72431 / 0.58853 \\
    PKUSUMSUM - LexRank & 0.39544 / 0.14940 & 0.58085 / 0.32428 \\
    PKUSUMSUM - TextRank & 0.41164 / 0.16862 & 0.64032 / 0.42380 \\
    PKUSUMSUM - Unsup. Submod. & 0.42699 / 0.18426 & 0.62063 / 0.41153 \\
    DocSumm - Greedy TFIDF & 0.39928 / 0.17318 & 0.66867 / 0.52939 \\
    NewsSumm - ILP with Budget & 0.41619 / 0.18641 & 0.62037 / 0.44923 \\
    NewsSumm - ScoreILP with TFIDF & 0.39358 / 0.16135 & 0.56496 / 0.36475\\
    NewsSumm - Title Reduction & 0.41717 / 0.17841 & 0.64739 / 0.48364 \\
    \hline

    \end{tabular}
    \label{tab:summarizers}
\end{table}

We see that all of the extractive summarizers are performing competitively against the ``Lead'' baseline. When comparing against human-annotated extractive summaries, we see two things. First, that summarizers are doing a good job capturing content that humans believe are important. Second, that when summarizers are compared against annotated extractive summaries they perform much better. From the small drop from R1 to R2 score of the Lead method, it is clear that most annotators' generated abstractive summaries also included more sentences from the beginning of the article. 

\subsection{Qualitative Evaluation of Extractive Summarizers}
Here we use word cloud representations to give an intuitive interpretation of the content in the generated extractive summarizers. We create word clouds for the two best methods in Section~\ref{ssec:eval-auto-summ}. In this paper, we used an online tool called WordItOut\footnote{\url{https://worditout.com}} to generate the word cloud representations. In all word clouds presented in this paper, a filter is used to display only the words with minimum frequency of 2.

\begin{figure}[!htb]
\centering
\includegraphics[width=\textwidth]{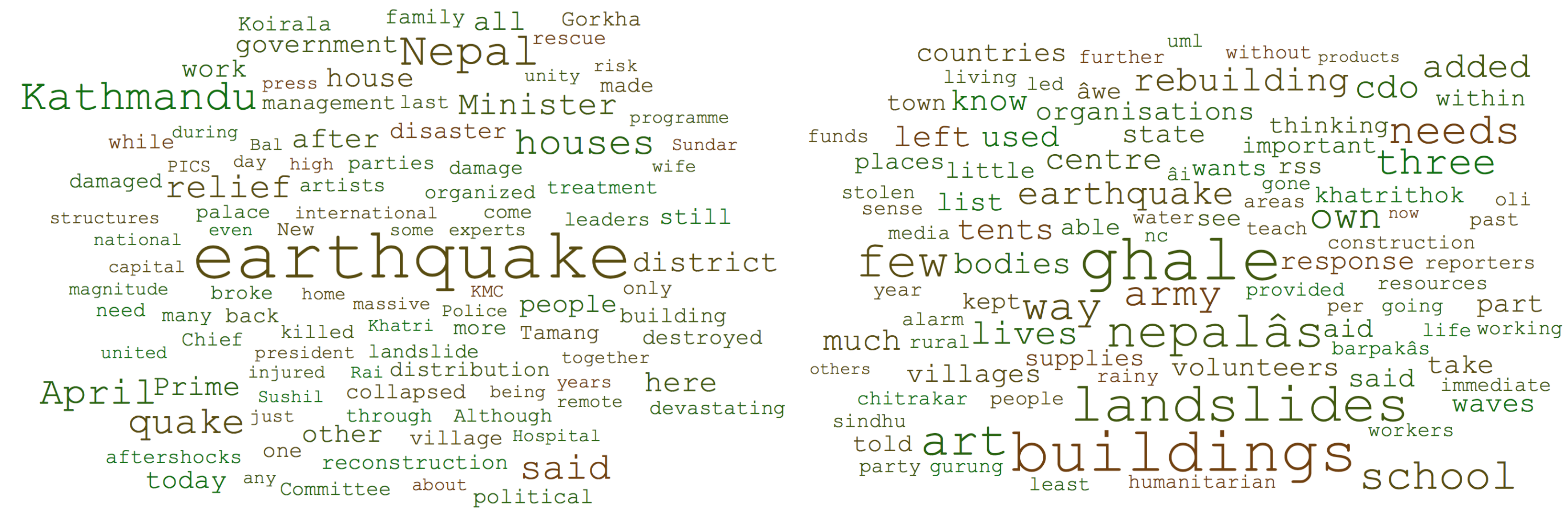}
\caption{The word clouds representing summaries generated by PKUSUMSUM-Centroid method (left) and original documents without the content of those summaries (right).}
\label{fig:CentriodSummDiff}
\end{figure}

Figure~\ref{fig:CentriodSummDiff} shows a word cloud made by the aggregation of all the summaries generated by the PKUSUMSUM-Centroid method. This gives a sense of the content in those summaries. For contrast, we also generate a word cloud for the original news articles without the content of the generated summaries. Specifically, common words are first removed completely and then the word clouds are built with frequencies of surviving words. In essence, this shows what information remains apart from the generated summaries.

\begin{figure}[!htb]
\centering
\includegraphics[width=\textwidth]{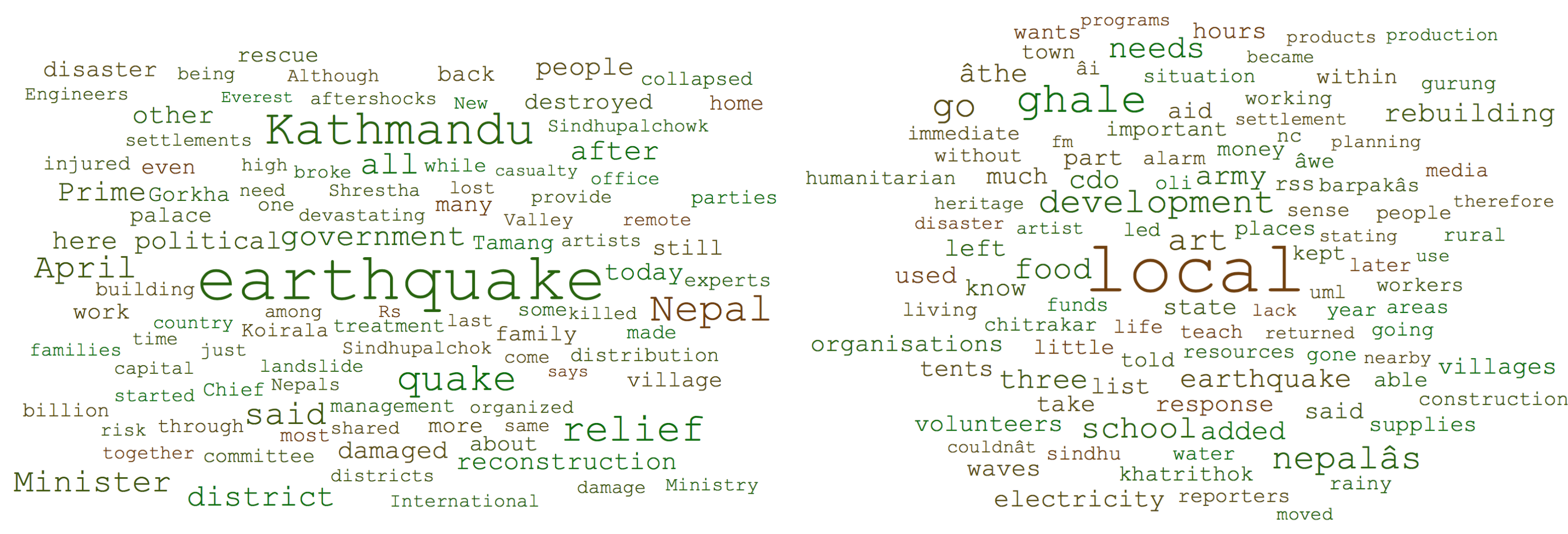}
\caption{The word clouds representing summaries generated by PKUSUMSUM-Lead method (left) and original documents without the content of those summaries (right).}
\label{fig:LeadSummDiff}
\end{figure}

The images clearly show a contrast of content. The summary wordcloud shows ``earthquake'' as its most prominent word. The image of the articles show less focus. If viewed alone, the reader would not quickly infer the gist of the original content. Similarly, Figure~\ref{fig:LeadSummDiff} represents ``Lead'' method. And here we see an even more stark difference.

\section{Comparing Twitter and News Media information}
\subsection{News Reporting Speed}
\label{NewsRepSpeedSec}
\begin{figure}[!htb]
\centering
\includegraphics[width=\textwidth]{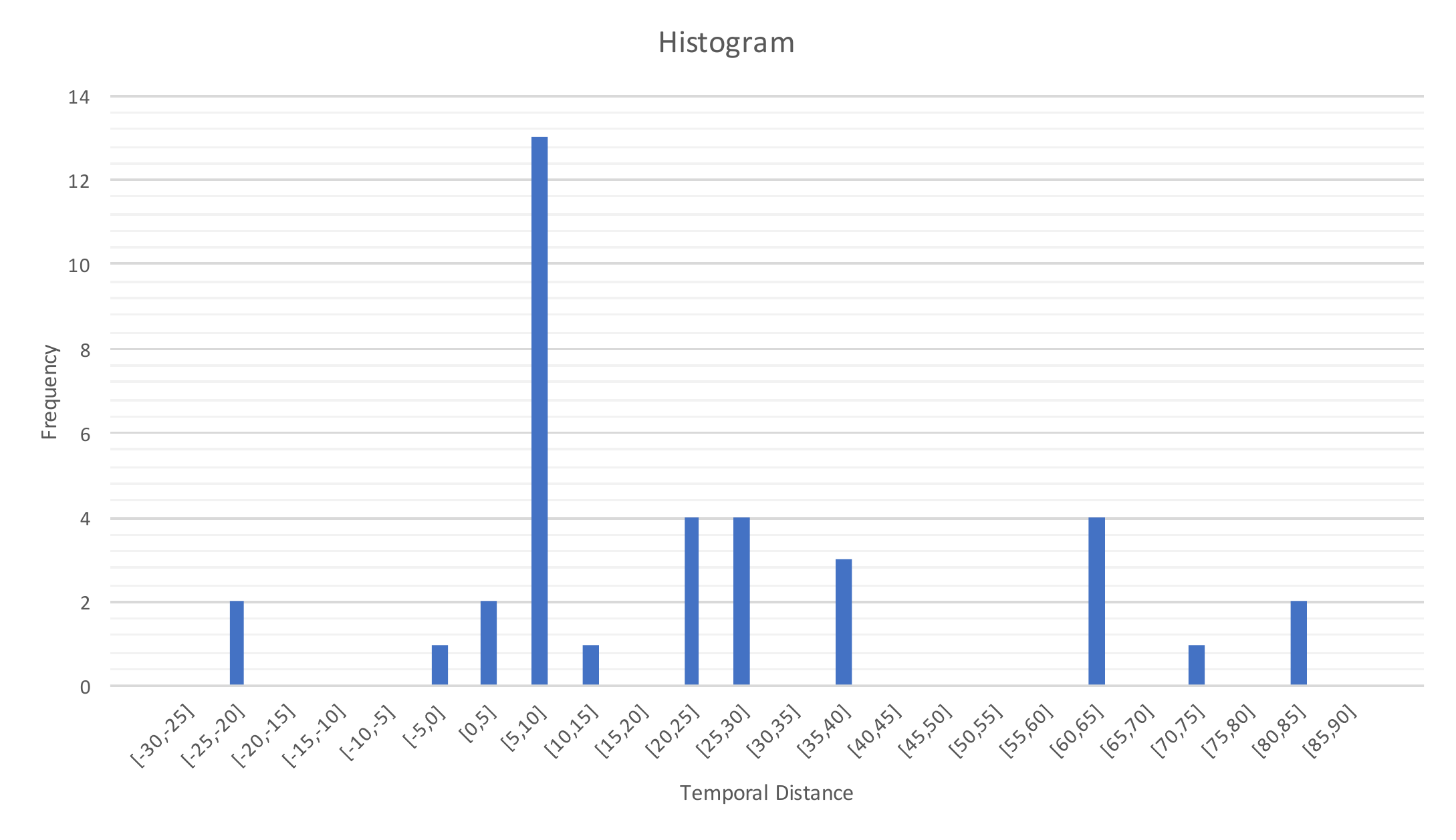}
\caption{The histogram of the temporal distances of the matched pairs annotated as relevant.}
\label{fig:RelTemporalDistanceHistogram}
\end{figure}

\begin{figure}[!htb]
\centering
\includegraphics[width=\textwidth]{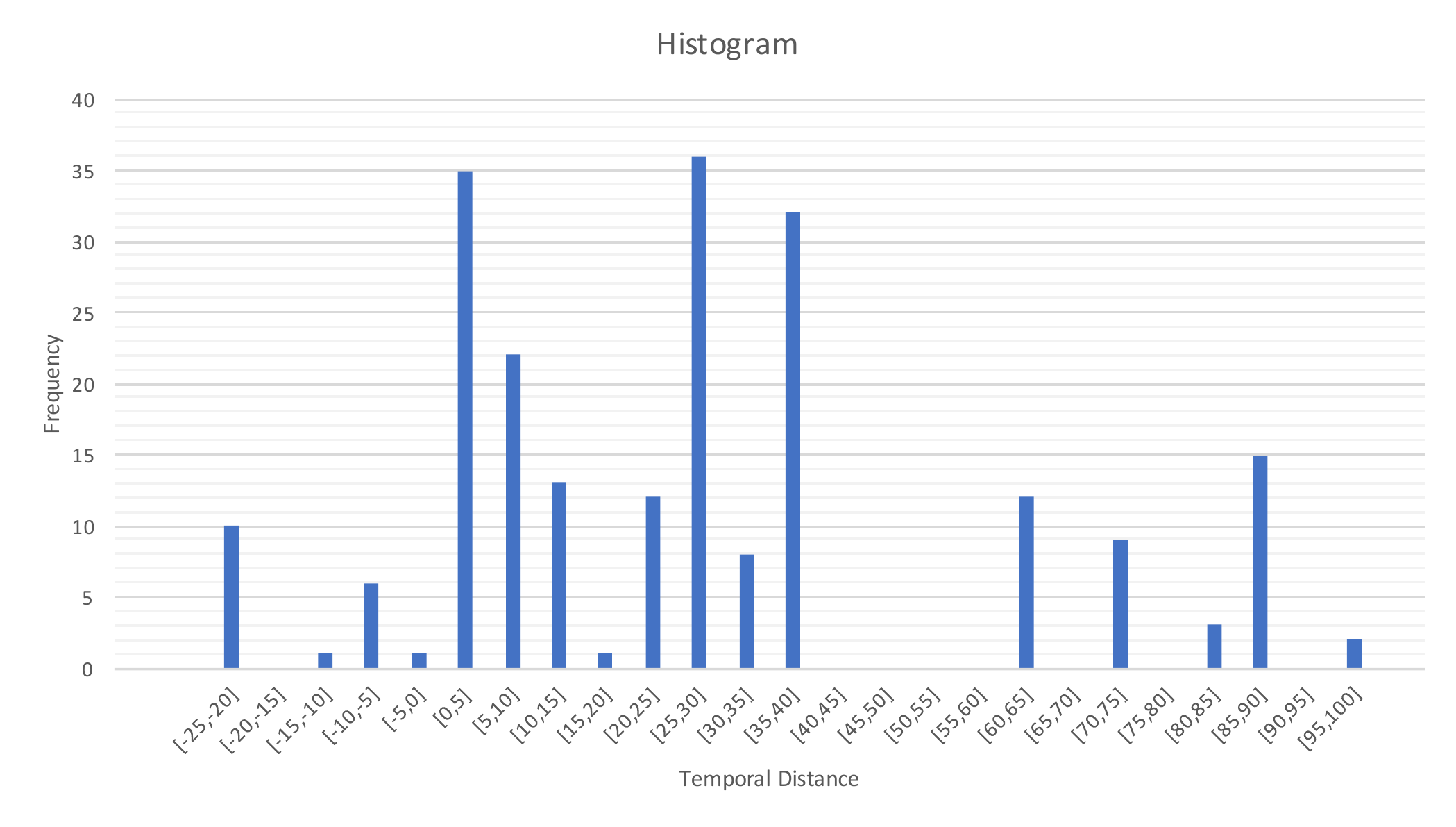}
\caption{The histogram of the temporal distances of the matched pairs annotated as partially relevant.}
\label{fig:PartiallyRelTemporalDistanceHistogram}
\end{figure}

One of the aspects of comparing Twitter and traditional newswire is their speed in reporting news, specially during a disaster. To this aim, we used the set of tweets-news article pairs classified as matched by the proposed method and also annotated as relevant or partially relevant pairs by annotators. For these pairs, we calculated the temporal distance between the tweets and their corresponding news article by subtracting the news article publish date from tweet publish date. We computed the temporal distances for all matched pairs annotated as relevant and partially relevant separately. Figures \ref{fig:RelTemporalDistanceHistogram} and \ref{fig:PartiallyRelTemporalDistanceHistogram} show the histograms of the temporal distances of the matched pairs annotated as relevant
and partially relevant by bin. The size of the bins is 5. 

As shown in Figures~\ref{fig:RelTemporalDistanceHistogram} and \ref{fig:PartiallyRelTemporalDistanceHistogram}, in most cases the temporal distances are positive that means in most matched pairs of tweet-news, the news article publish date is older than its matched tweet date. Furthermore, the percentage of tweets that appeared before their matching news article are 91.8\%  and 91.7\% in relevant and partially relevant pairs, respectively. This implies that the news are reported by tweets faster than news articles in both relevant and partially relevant pairs.

\subsection{Clustering on Tweets} \label{tweet-events}
To investigate the second research question (RQ2), we used the same set of tweets that was used in \ref{NewsRepSpeedSec} to compare
the news reporting speed in Twitter and news media, then we removed the few number of tweets that were published after their matched news articles and employed the resulting tweet set for clustering. In other words, we obtained a set of tweet-news pairs that were classified as matched by our tweet-news linking method, annotated as relevant or partially relevant by annotators and were published before
their matched news article to find that what type of information are reported {\em earlier} by Twitter through investigating the clusters found on the tweets. 

For clustering, we used a short text clustering method called GSDMM which is a collapsed Gibbs Sampling algorithm for the Dirichlet Multinomial Mixture model \cite{yin2014dirichlet}. 
In this experiment, the values of the GSDMM\textsc{\char13}s parameters are $k = 15$, $\alpha= 0.1$, $\beta= 0.1$, $t= 50$ where $k$ is the number of clusters, $\alpha$ and $\beta$ are Dirichlet priors and $t$ is the number of iterations that the clustering algorithm repeats until convergence. The total number of tweets used as input in this experiment is 172. Table \ref{tab:eventDetectionRes} contains some statistics about the clustering result. In this experiment, tweets are preprocessed by removing the non-ascii characters, punctuation, stop-words and URLs.

\begin{table}[h]
    \centering
    \caption{Some statistics about the clustering results on tweets.}
    \begin{tabular}{cccc}
        \hline
        Min & Max & Average & Mode of \\
        cluster size &  cluster size &  Cluster size & the cluster sizes \\ \hline
        8 & 18 & 11.47 & 9 \\ \hline
    \end{tabular}
    \label{tab:eventDetectionRes}
\end{table}

\begin{figure}[!htb]
\centering
\includegraphics[width=\textwidth]{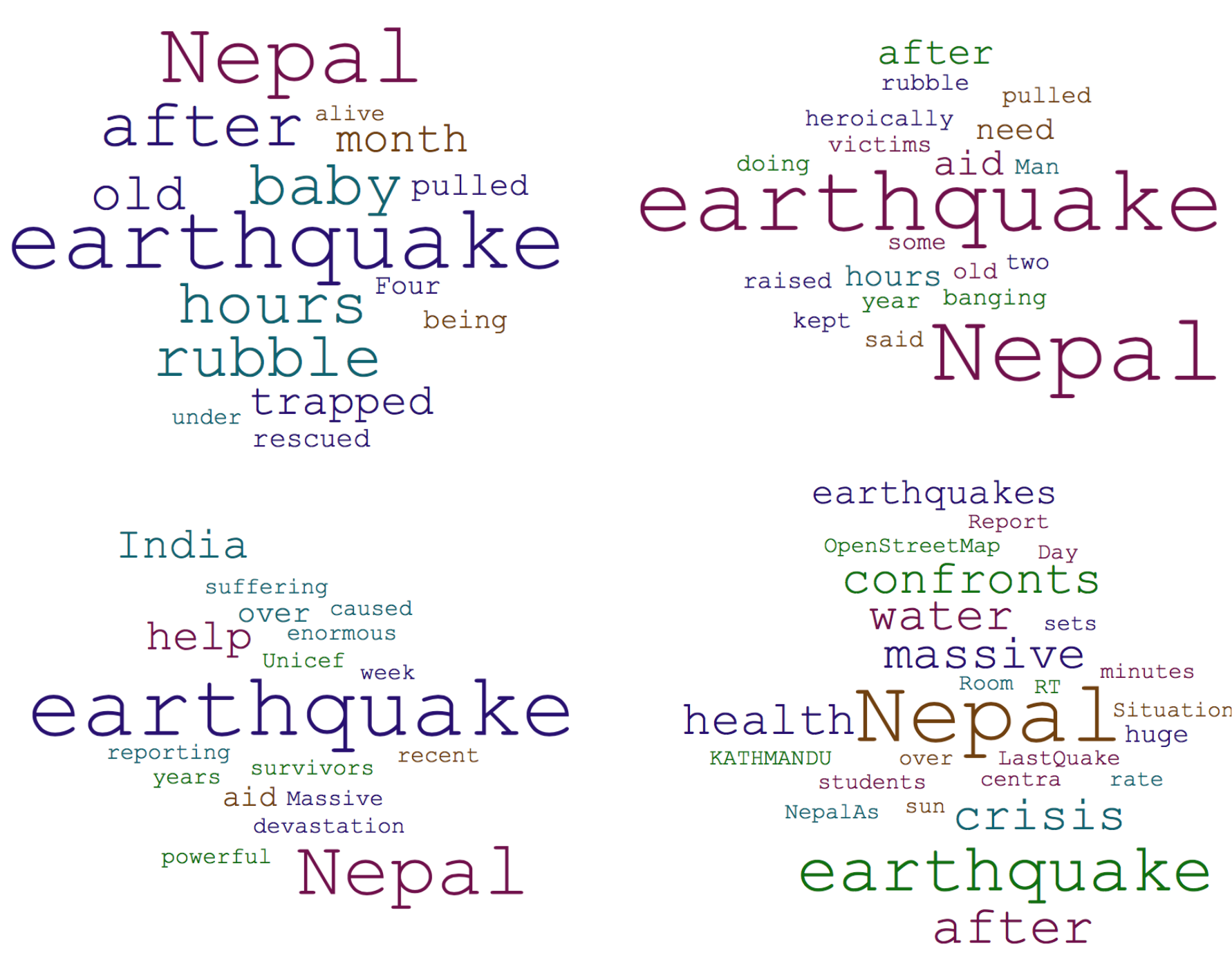}
\caption{The events obtained by GSDMM method using the tweets that appeared before their corresponding news articles.}
\label{fig:events}
\end{figure}

Some of the clusters obtained from applying GSDMM on the set of tweets mentioned above are shown in Figure~\ref{fig:events}. 
The word cloud representation is used to show the clusters. As Figure~\ref{fig:events} shows, each of these clusters correspond to a separate topic relevant to the Nepal Earthquake. The topics corresponding to each of these clusters include a four month old baby being pulled form the  rubble (the top left cluster), a man being pulled from the rubble 82 hours after the Nepal earthquake (the top right cluster), India's effort to help the earthquake survivors (the bottom left cluster) and the health water crisis in Nepal after the earthquake (the bottom right cluster). 

\subsection{Content Comparison}
\begin{figure}[!htb]
\centering
\includegraphics[width=\textwidth]{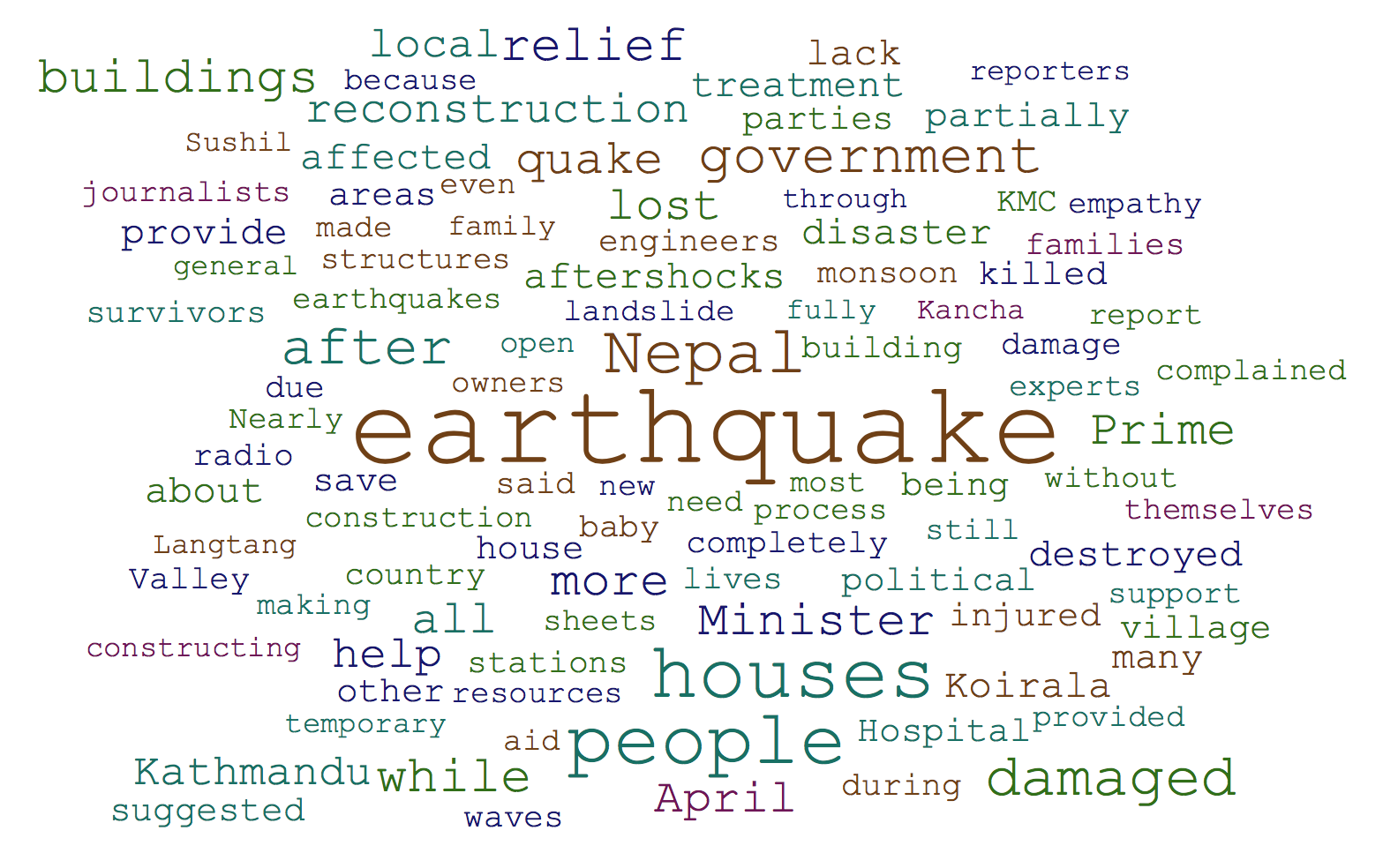}
\caption{The word cloud representation of the news articles' human annotated summaries.}
\label{fig:SummWC}
\end{figure}

\begin{figure}[!htb]
\centering
\includegraphics[width=\textwidth]{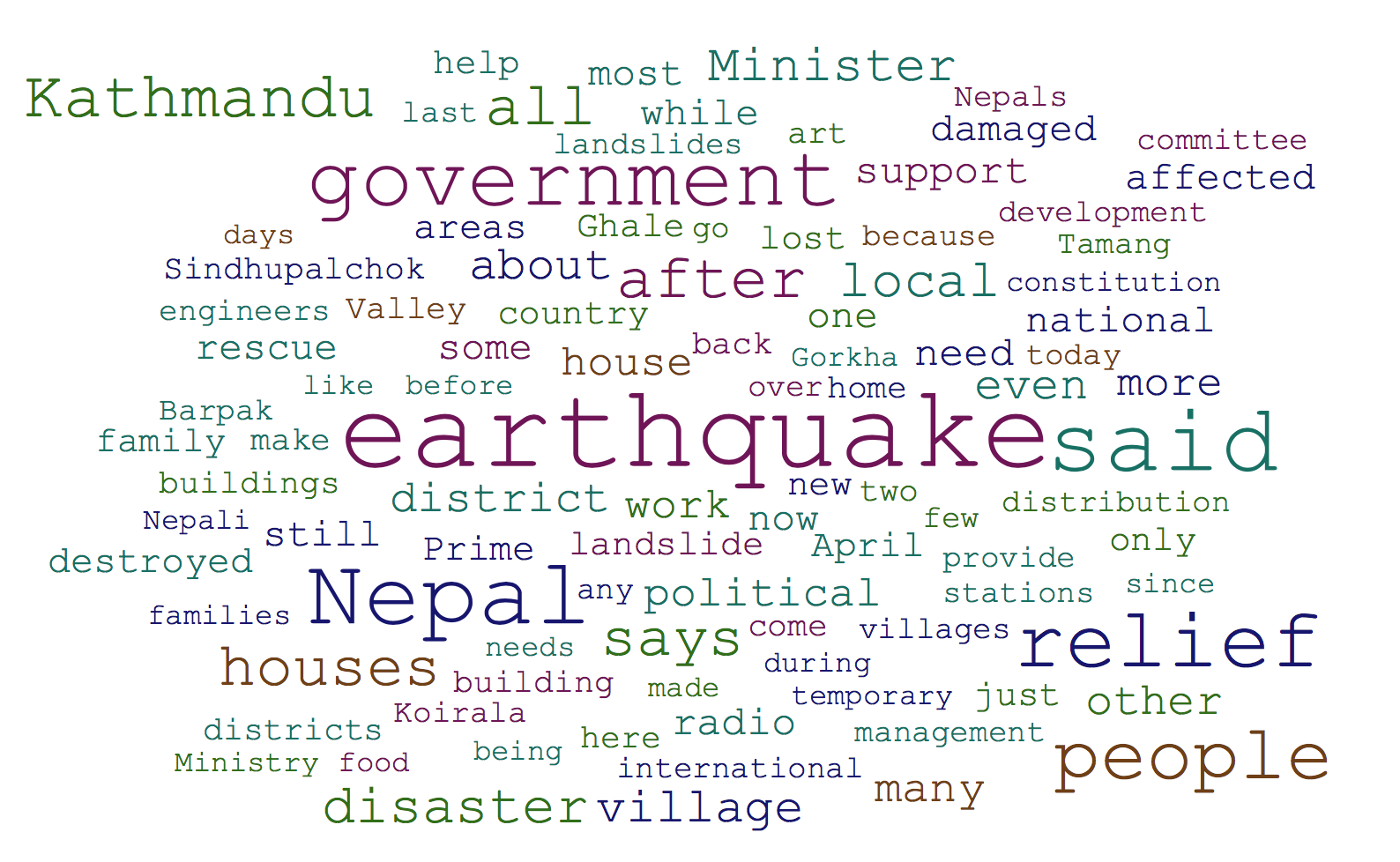}
\caption{The word cloud representation of the news articles'  content.}
\label{fig:NewsWC}
\end{figure}

\begin{figure}[!htb]
\centering
\includegraphics[width=\textwidth]{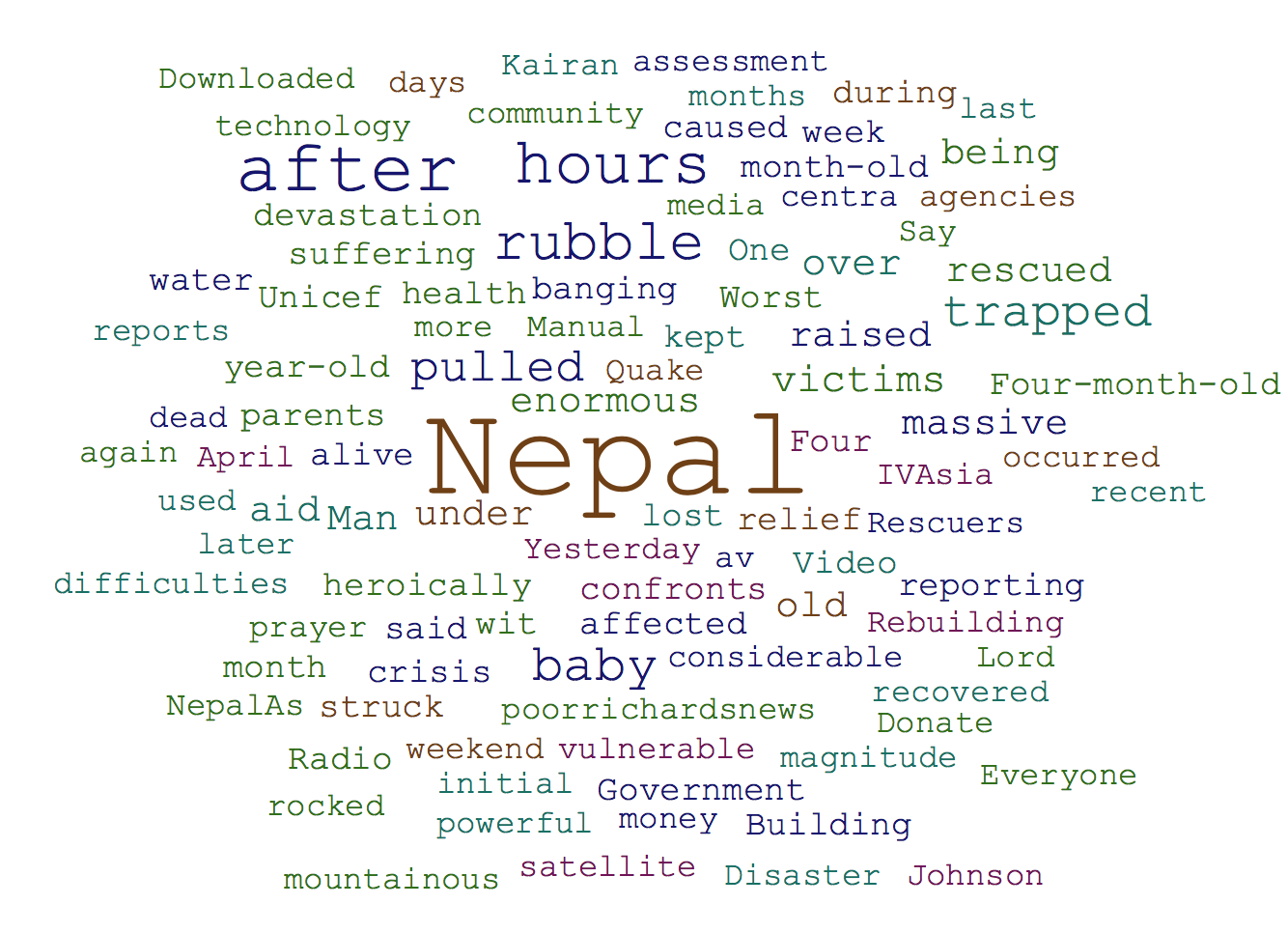}
\caption{The word cloud representation of the tweets' content that were annotated as relevant.}
\label{fig:RelTWC}
\end{figure}

\begin{figure}[!htb]
\centering
\includegraphics[width=\textwidth]{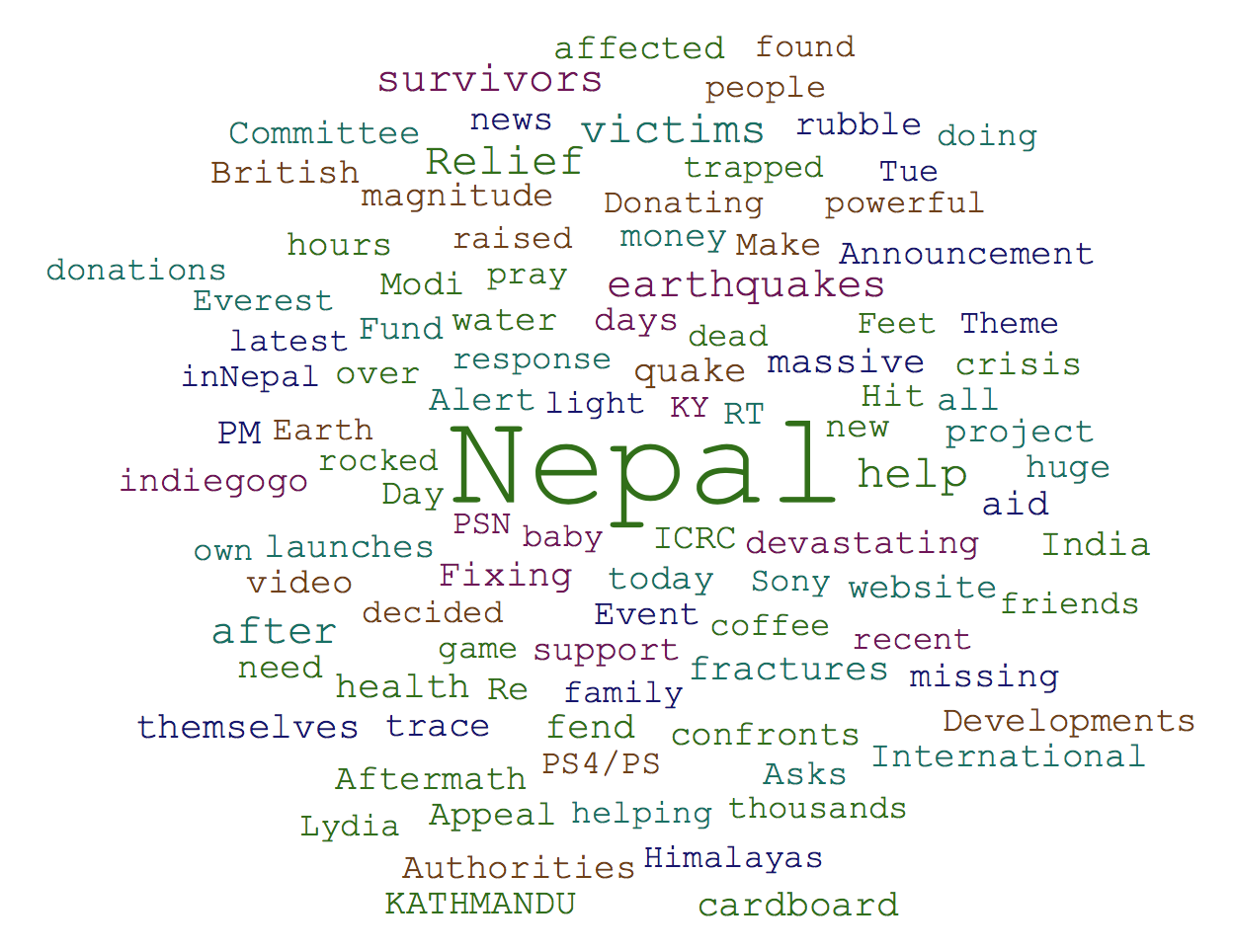}
\caption{The word cloud representation of the tweets' content that were annotated as partially relevant.}
\label{fig:ParRelTWC}
\end{figure}

In this section, the word cloud representation is used to compare the textual content of news articles, their annotated summaries and also the tweets content for  relevant and partially relevant annotated pairs as shown in figures \ref{fig:SummWC}, \ref{fig:NewsWC}, \ref{fig:RelTWC} and \ref{fig:ParRelTWC}.

One of the differences of news articles
and tweets content, shown in their word clouds is the different choice of words in the two channels. As figures \ref{fig:NewsWC} and \ref{fig:SummWC} show objective words such as `government,' `people,' `houses' and `buildings' have high weights in news articles and their summaries, while as shown in figures \ref{fig:RelTWC} and \ref{fig:ParRelTWC}, more opinionated and subjective words such as `devastating,' `massive' and `suffering' are used in tweets.  Another observation is that the newspaper articles have a lot of locations, e.g., `Kathmandu,' `Sindhupalchok,' `Tamang,' and  `Barpak,' but the tweets focus more on the human angle, e.g., `parents,' `baby,' `man,' `month-old,' `four-month-old,' `year-old,' `trapped,' `pulled,' and `rubble.' 

\subsection{Tweets as Summaries}
The tweets deemed ``relevant'' can be grouped together and considered a summary. The relevancy was computed as explained in Section~\ref{sssec:tweet-relevance}. Using these tweet summaries, we again use ROUGE to evaluate their content. Table~\ref{tab:tweet-summ} shows a weak representation of news articles. However, the precision of the tweets against extractive annotations shows that tweets are  capturing pertinent information, within their tight size constraint.

\begin{table}[h]
    \centering
    \caption{Evaluating Tweets as Summaries}
    \begin{tabular}{lccc}
    \hline
    Relevancy & R1/R2 F1 & R1/R2 Prec. & R1/R2 Prec. \\
    Type & Abstractive & Extractive & Whole Article \\
    \hline
        Partially Relevant & 0.17021 / 0.01749 & 0.22569 / 0.02191 & 0.39487 / 0.05062 \\
        Relevant & 0.11670 / 0.01150 & 0.36622 / 0.02676 & 0.51154 / 0.06385\\
    \hline
    \end{tabular}
    \label{tab:tweet-summ}
\end{table}

Another approach to evaluating the content of these tweet summaries is to look at the ``unique'' information found only in the tweet summaries. If we view the words of tweet summaries as one set, $A$, and represent the target comparison as another set of words, $B$, the we can use the following to show ``unique'' words found: 
\begin{equation}
    \label{Q:A-uniqueness}
    Uniq(A|B) = \frac{A-(A\cap B)}{A}
\end{equation}
\begin{equation}
    \label{Q:B-uniqueness}
    Uniq(B|A) = \frac{B-(A\cap B)}{B}
\end{equation}
Table~\ref{tab:tweet2article} show that as expected the news articles themselves contain a lot of words not in the tweet summaries (93.4\% and 96.9\% for partial relevance and full relevance, respectively). But it also shows that tweet summaries contain unique words as well. This, in conjunction with the good precision values, reveals that the tweet summaries do contain some pertinent and some unique information.

\begin{table}[h]
    \centering
    \caption{Uniqueness test of partially relevant tweet summaries (PRT) and relevant tweet summaries (RT) against news articles (NA), abstractive summaries (AS) and extractive summaries (ES)}
    \begin{tabular}{lccc}
        \hline
        Set $A$ & Set $B$ & $Uniq(A|B)$ & $Uniq(B|A)$ \\
        \hline
        PRT & NA & 0.76239 & 0.93424 \\
        RT & NA & 0.66994 & 0.96928 \\
        PRT & AS & 0.88186 & 0.90919 \\
        RT & AS & 0.81098 & 0.94268 \\
        PRT & ES & 0.85351 & 0.92084 \\
        RT & ES & 0.74424 & 0.95237 \\
        \hline
    \end{tabular}
    \label{tab:tweet2article}
\end{table}

From both Tables \ref{tab:tweet-summ} and  \ref{tab:tweet2article}, we see that the human annotators are using their own words to describe the events and thus the overlap is lower and the difference is higher with respect to abstractive summaries. The relevant tweets have the highest precision with the whole news articles indicating that they meaningful precursors of events/information described in more detail in the articles. However, the tables clearly show a complementary situation, i.e., the tweets relevant or partially relevant have some but not all of the information in the news articles and vice versa.

\subsection{The Extra Information in Tweets vs News}
In our next investigation, we tried to find what new information is provided in tweets that is not available in news media about the Nepal Earthquake and vice versa, i.e. what new information is provided in news articles that is not available in tweets. In this experiment, we used the same 172 tweets and news articles used in Section \ref{tweet-events}. To this aim, we computed the difference of the aggregate of all tweets' content, denoted by $T_{172}$, from the aggregate of the matching news articles' content (NAC), i.e. $NAC - T_{172}$, and $T_{172} - NAC$, and represented the resulting set of words by  their word clouds. 
Precisely speaking, to compute $NAC - T_{172}$, we find the common words between $NAC$ and $T_{172}$ and remove all of its occurrences from $NAC$ . We do the same for computing $NAC - T_{172}$.

\begin{figure}[!htb]
\centering
\includegraphics[width=\textwidth]{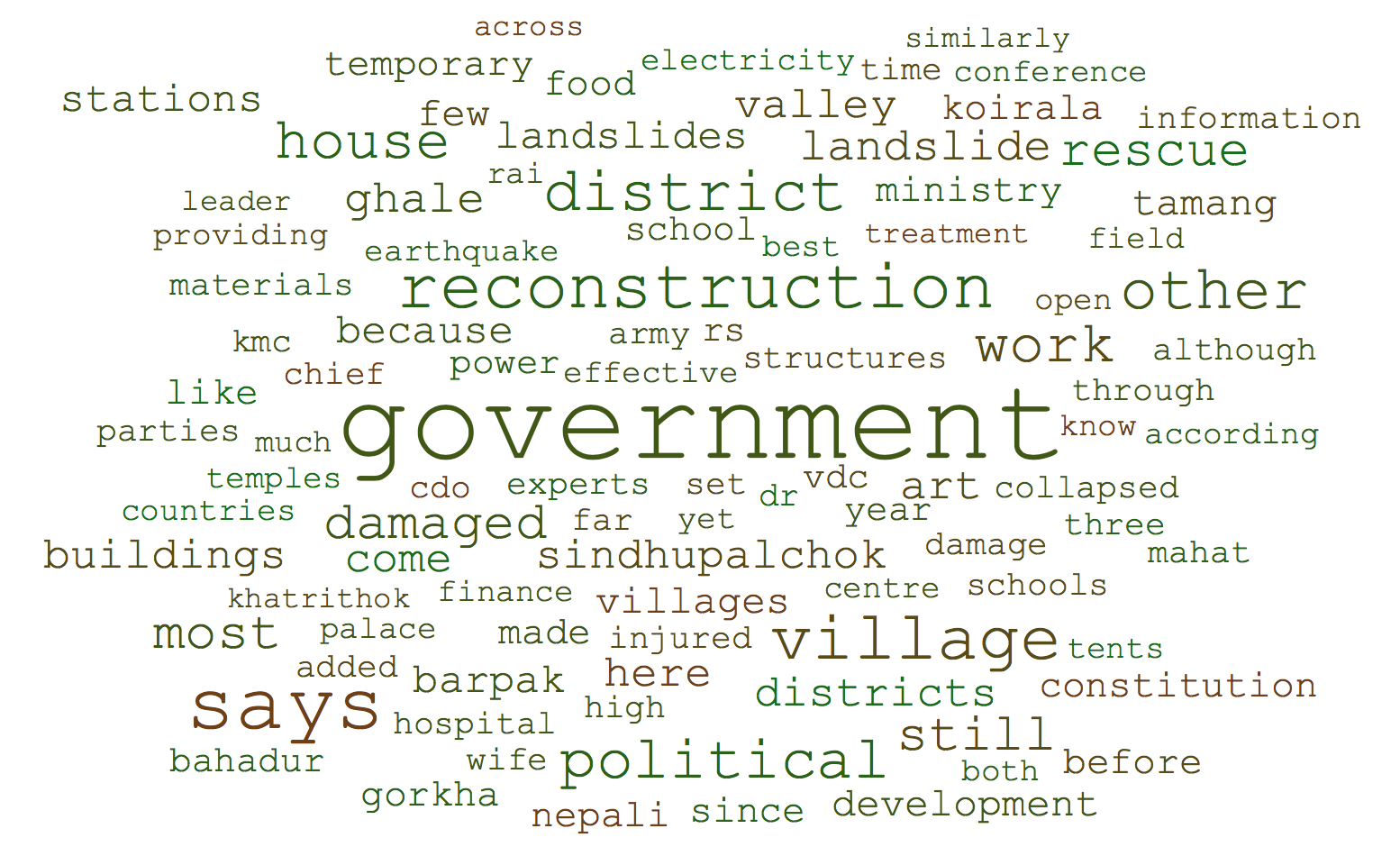}
\caption{The word cloud representation of the news set of words subtracted by the tweets set of words.}
\label{fig:NewsMinusTweets}
\end{figure}

\begin{figure}[!htb]
\centering
\includegraphics[width=\textwidth]{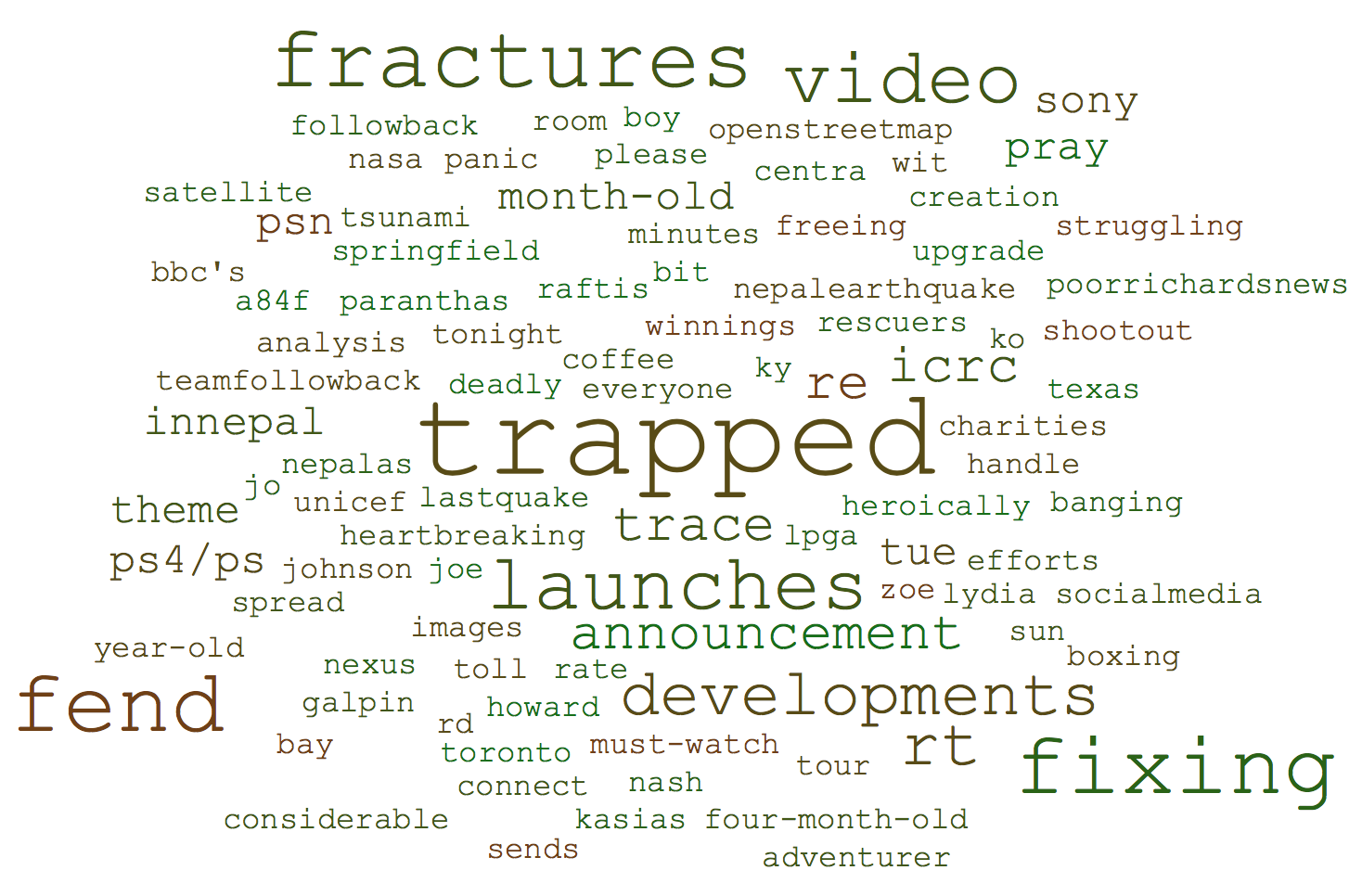}
\caption{The word cloud representation of the tweets set of words subtracted by the news set of words.}
\label{fig:TweetsMinusNews}
\end{figure}

 Comparing the two word clouds in Figures \ref{fig:NewsMinusTweets} and \ref{fig:TweetsMinusNews} shows that news articles mainly contain information about the Nepal earthquake from the agencies viewpoint. The use of terms such as `government', `reconstruction' and `political' shows the formal and objective language employed by news agencies to describe the situation.
 In contrast, tweets content is more about the human angle, the content type that news articles lack. These word clouds reinforces that the two sources of information can complement each other to provide a comprehensive picture from two different angles about a major disaster like the Nepal Earthquake. The immediacy of tweets can be used by first responders for search and rescue operations. 

\section{Conclusion} \label{conclusion}
In this paper, we studied, compared and analyzed newswire and Twitter from different viewpoints in the context of the 2015 Nepal Earthquakes. In this regard, we collected and annotated two datasets: A tweet dataset that contains 336,140 tweets related to the Nepal Earthquake written from April 24, 2015 to June 25, 2015 and a news dataset containing 700 news articles relevant to the Nepal Earthquakes and dated in the year 2015. We presented descriptive statistics of the collected tweets from different viewpoints:  the most frequent words, the top most active users, the changes in the number of tweets over time and the use of hashtags, URLs and mentions in the collected tweets.  

We also evaluated several methods of summarization of news articles against human ground truth and against  Twitter content. Furthermore, using the tweet-news pairs classified as matched and annotated as relevant, we compared the speed of Twitter and newswire in news reporting (RQ1), the content generated in each of these two channels (RQ3, RQ4), and the effectiveness of the matched tweets to summarize their corresponding news articles (RQ6). We found that during the Nepal earthquake, most of the human news and earthquake related crises appear in Twitter {\em before} news media. Another finding is that during a major disaster, twitter contains more opinionated and subjective content in comparison with the news media content.

We have also shown that automatic summarization can be an effective method for quickly pulling the important information from news articles. All the methods investigated in this paper were {\em unsupervised} approaches, which means that they can be used as quick and efficient filters to combat the information overload present in news articles. 
We also show that twitter data holds data that is complementary to the content of relevant news articles. For consumers of information like first responders, it is paramount that all available information about a natural disaster can be quickly processed. Both automatic summarization of news articles as well as Twitter content, can be used to support this need.
We also proposed a tweet-news linking method to find the matched tweets to each news article and evaluated its performance using our annotated datasets (RQ5). It gave a decent precision of $0.47$ on the annotated subset.


\section*{References}
\bibliographystyle{elsarticle-num}
\bibliography{main}

\end{document}